\newif\ifshowfig
\def\ifshowfig{\iftrue}

\newif\ifsubmit
\def\ifsubmit{\iftrue}
\documentclass[aps,pre,twocolumn,superscriptaddress]{revtex4-1}

\usepackage{graphicx}
\usepackage{bm,bbm}
\usepackage{amssymb,amsfonts,amsmath}
\usepackage[english]{babel}
\usepackage[utf8]{inputenc}
\usepackage{xcolor}
\usepackage{hyperref}
\usepackage{ulem} \renewcommand{\emph}[1]{\textit{#1}}

\DeclareMathOperator{\argmin}{arg\,min}
\begin{document}

\title{Neural network approach to time-dependent dividing surfaces in 
classical reaction dynamics}
\author{Philippe Schraft}
\author{Andrej Junginger}
\thanks{Present address: Machine Learning Team at ETAS GmbH (Bosch Group).}
\author{Matthias Feldmaier}
\author{Robin Bardakcioglu}
\author{J\"org Main}
\author{G\"unter Wunner}
\affiliation{Institut f\"ur Theoretische Physik 1, Universit\"at Stuttgart, 
  70550 Stuttgart, Germany}

\author{Rigoberto Hernandez}
\email{Correspondence to r.hernandez@jhu.edu}
\affiliation{Department of Chemistry,
The Johns Hopkins University,
Baltimore, MD, USA}

\date{\today}

\newcommand{\ie}{i.\,e.}
\newcommand{\eg}{e.\,g.}
\newcommand{\cf}{cf.}
\newcommand{\etal}{\textsl{et~al.}}
\newcommand{\FIG}{Fig.}
\newcommand{\FIGS}{Figs.}
\newcommand{\SEC}{Sec.} \newcommand{\SECS}{Secs.}
\newcommand{\EQ}{Eq.}
\newcommand{\EQS}{Eqs.}
\newcommand{\REF}{Ref.}
\newcommand{\REFS}{Refs.}
\newcommand{\comment}[1]{\textcolor{orange}{\textsf{#1}}}   
\definecolor{mygreen}{RGB}{0,200,0} 
\newcommand{\Ng}{{N_\text{g}}}
\newcommand{\Ns}{{N_\text{s}}}
\newcommand{\transpose}{\mathsf{T}}
\newcommand{\xx}{\vec{x}}
\newcommand{\vv}{\vec{v}}
\newcommand{\prest}{\bar{p}}
\newcommand{\ue}{\mathrm{e}}
\newcommand{\ud}{\mathrm{d}}
\definecolor{myred}{RGB}{200,0,0}
\definecolor{mygreen}{RGB}{0,150,0}
\renewcommand{\vec}[1]{\bm{#1}}
\newcommand{\matr}[1]{\mathbf{#1}}
\newcommand{\subbath}{_\text{bath}}
\newcommand{\subreac}{_\text{reac}}
\newcommand{\sigmoid}{\sigma}
\newcommand{\Ws}{\mathcal{W}_\text{s}}
\newcommand{\Wu}{\mathcal{W}_\text{u}}
\newcommand{\Wsu}{\mathcal{W}_\text{s,u}}
\newcommand{\LD}{\mathcal{L}}
\newcommand{\TSt}{\mathcal{T}}

\hyphenation{mani-fold}
\hyphenation{mani-folds}


\begin{abstract}
In a dynamical system, the transition between reactants and products is 
typically mediated by an energy barrier whose properties determine the 
corresponding pathways and rates.
The latter is the flux through a dividing surface (DS) between 
the two corresponding regions and it is exact only if it is free of recrossings.
For time-independent barriers, the DS can be attached to the top of the 
corresponding saddle point of the potential energy surface, and in 
time-dependent systems, the DS is a moving object.
The precise determination of 
these direct reaction rates, \eg~using transition state theory, 
requires the actual construction of a DS for a given saddle geometry which 
is in general a demanding methodical and computational task, especially in 
high-dimensional systems.
In this paper, we demonstrate how such time-dependent, global, and 
recrossing-free DSs can be constructed 
using neural networks.
In our approach, the neural network uses the bath coordinates and time as input 
and it is trained in a way that its output provides the position of the 
DS along the reaction coordinate.
An advantage of this procedure is that, once the neural network is trained, the 
complete information about the dynamical phase space separation is stored in the 
network's parameters, and a precise distinction between reactants and products 
can be made for all possible system configurations, all times, and with little 
computational effort.
We demonstrate this general method for two- and three-dimensional 
systems, and explain its straightforward extension to even more degrees of 
freedom.
\end{abstract}

\maketitle


\section{Introduction}

One of the grand challenges in reaction dynamics is the 
accurate determination of reaction pathways and rates.
Theoretical and computational approaches require both a 
precise understanding of the underlying potential energy landscape describing 
the reactive system and the effect of possible external
---that is, time-dependent--- forces.
Formally, all that then remains is the integration of the equations of motion
for a sufficiently large ensemble of trajectories and long times to completely
characterize the dynamics.
In practice, however, the dimensionality of reactive systems is sufficiently 
large that such approaches are computationally prohibitive. 
The key to resolving this challenge has long been known to lie in the observation that
reactions are typically mediated by an 
energy barrier separating reactant and product basins in the underlying 
state space.

Transition state theory (TST)~\cite{pitzer,pechukas1981,truh79,truh85,truhlar91,truh96,truh2000, 
Komatsuzaki2001,Waalkens2008,hern08d,Komatsuzaki2010,hern10a,Henkelman2016} 
has been a powerful approach for obtaining chemical reactions rates
approximately.
TST rates are determined by the flux through a dividing surface (DS) 
separating reactants and products divided by the reactant population.
These rates are exact if and only if the DS is free of 
recrossings, \ie~it is crossed by each reactive trajectory exactly once.
It should be noted that the boundary conditions
imposed on the reaction process 
are taken to be such that the particles that reach the reactant or product
basins never return either because the potential is unbound,
geminate recombination is quenched, or particles are not propagated
beyond these basins.
For the overall rate, one can sometimes combine the direct rates 
between basins as was recently done in Refs.~\cite{hern16c,hern17e}.
Many of the advances over the past century have hinged on the 
construction of DSs with decreasing degree of recrossing.
Thus, a crucial step to calculate accurate TST rates is the precise 
determination of a nonrecrossing hypersurface, 
and this is a nontrivial task especially in 
high-dimensional and time-dependent 
systems.

In autonomous systems, a \emph{local} DS can be constructed using the 
normally hyperbolic invariant manifold~\cite{pollak78,pech79a,hern93b,hern94,Uzer02,Teramoto11,Li06prl,%
Waalkens04b,Waalkens13}, \eg~using normal form transformations~\cite{Murdock02}.
However, the situation quickly becomes daunting when the system is 
time-dependent or if one aims to obtain a \emph{global} DS (at points far
from the saddle region.)
In the former case, the DS itself is a time-dependent object that is 
harder to calculate, whereas in the latter case, perturbative approaches
break down because of a finite radius of convergence.

For systems with one and two degrees of freedom this problem can basically be solved
using a general minimization procedure based on Lagrangian descriptors (LDs)~\cite{Mancho2010,Mancho2013}.
In the one-dimensional case, the moving DS is attached to the time-dependent 
transition state trajectory~\cite{dawn05b,
	dawn05a,
	hern06d,
	hern14b,hern14f,hern15a,Kawai2009a}.
If additional bath degrees of freedom are taken into account, 
these approaches 
are formally extendable to 
higher dimensions, and some of us have demonstrated this
with at least one additional bath mode~\cite{hern17h}.
Its structure is directly related to special unstable  
trajectories close to the saddle
which are challenging to construct even in the one-dimensional case.
Additional bath degrees of freedom
lead to increasingly chaotic trajectories 
and increase the dimensionality of the time-dependent DS, 
making it increasingly challenging to obtain a single point of the DS.
As such a process leads only to a finite number of points, 
one remains with the additional challenge of obtaining
a continuous surface.

In this paper, we demonstrate how all these issues become manageable 
when neural networks (NNs) are trained to approximate the DS.
Although NNs have been known for decades, they have regained 
intense attention over the recent years, mostly because of their large 
success in the fields of image recognition and classification
and the increasing power of computers to handle large datasets
and complex representations. 
Recently, Carpenter et al.~\cite{wiggins17} demonstrated the use of 
such machine learning techniques 
to obtain molecular dynamics trajectories from which one can extract phase space
structures.
Further, NNs and other machine learning methods are also used to construct high-dimensional potential energy surfaces and other quantities in theoretical chemistry~\cite{blank1995neural,behler2007generalized,behler2011atom,cui2016efficient,vargas2017machine,
rupp2012fast,cui2015gaussian_a,cui2015gaussian_b,faber2016machine,huang2017chemical}.
Here we show that we can construct one such phase space structure
---the DS--- directly using a NN.
Our method is based on the fact that \emph{single} points on the DS can be
calculated with reasonable effort as well as easily propagated in time:
They are easy to obtain, because they are minima of the LD for a given time 
$t$ and, in addition, their dynamics is readily obtained by numerically integrating 
the equations of motion.
To obtain the DS over the entire domain, we train a NN on a set of such points.
The input to the NN is the system's configuration and 
its output is the position of 
the DS along the reaction coordinate.
From the machine learning perspective, what we present in this paper is a 
classical, multidimensional regression task that we solve using feed-forward 
NNs.

This paper is organized as follows:
In \SEC~\ref{sec:theory}, we review the basic concepts of calculating 
nonperturbative DSs by minimizing LDs, and that of NNs.
In \SEC~\ref{sec:results}, we present the application of the method to 
two- and three-dimensional reactive systems.

\section{Theory}
\label{sec:theory}

\subsection{Time-dependent dividing surfaces in classical reaction dynamics}
\label{sec:theory:LDs}
\begin{figure}
\includegraphics[width=\columnwidth]{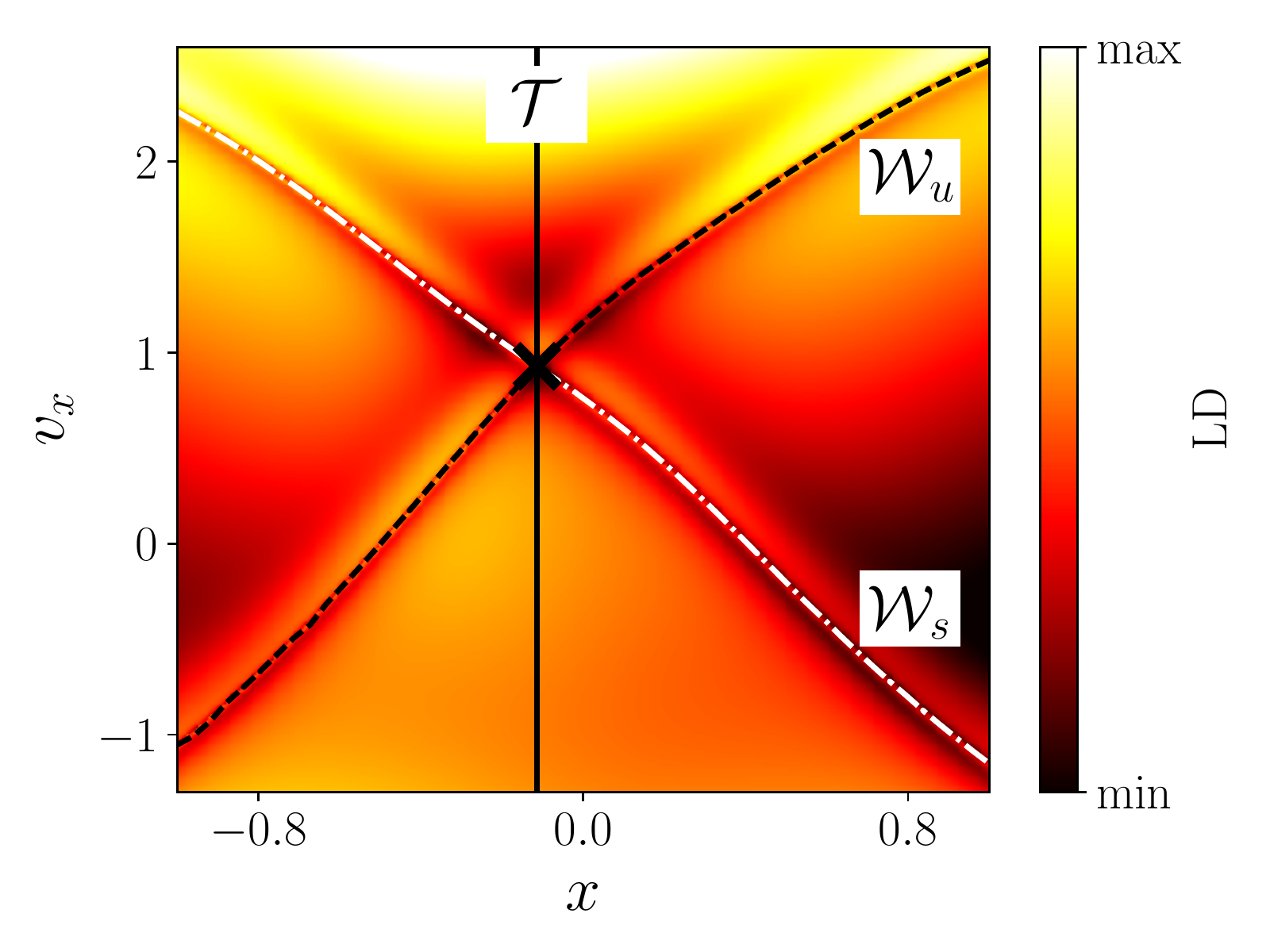}
\caption{
Plot of the LD~\eqref{eq:LD} of the rank-1 saddle defined by the 
potential~\eqref{eq:model_pot} for $y=z=v_y=v_z=t=0$.
The axes approximate the reaction coordinate $x_\mathrm{reac}$ and its associated velocity.
The color coding represents the value of the LDs.
The stable and unstable manifolds $\Wsu$ are visible as local minima of the LD.
The stable manifold $\Ws$ is highlighted by the white dash-dotted curve and the 
unstable manifold $\Wu$ by the black dashed one.
Their intersection is marked with a black cross ($\times$) and denotes one 
point of the NHIM of the corresponding saddle.
The solid vertical black line is the DS, $\tau$, attached to this point of the NHIM.
The choice of the DS is, in general, not unique.
We chose the most simple case, viz.\ a vertical line independent of $v_x$,
where we assume that the DS does not cross any of the manifolds in an
additional point.
Such an intersection exists for each configuration of bath coordinates and for 
each time.
See text for further explanations.
}
\label{fig:sketch-of-intersection}
\end{figure}

In dynamical systems where reactants and products are separated by an energetic 
barrier, each reactive particle has to overcome the 
barrier region to reach
the corresponding opposite basin, \eg~the product basin and reactant basin,
respectively.
In accordance with TST,
the local properties of the barrier determine both the reaction pathways
and the corresponding rates.
The stable and unstable manifolds
$\Wsu$ play an important role in this picture.
Here, the stable manifold $\Ws$ is the set of points that exponentially 
approach 
a time-dependent normally hyperbolic invariant manifold
(NHIM) and the unstable manifold $\Wu$ are those points
that exponentially depart from the NHIM.
These manifolds have a crucial meaning
because they separate reactive from
non-reactive trajectories (see \FIG~\ref{fig:sketch-of-intersection}).
The intersection of 
the closure of these manifolds
---here meant in a geometric, not a dynamic sense---
is the NHIM to which a recrossing-free DS
can be attached that separates the reactant from the
product basin.

The lowest-energy pathway crosses the barrier region at a rank-1 saddle point which 
locally defines a set of $d-1$ bath coordinates and 
a one-dimensional reaction coordinate 
that are parallel and perpendicular, respectively, to the associated 
$(2d-2)$-dimensional NHIM.
The attached DS has an increased dimension $(2d-1)$
in respect to the NHIM, as can be seen in \FIG~\ref{fig:sketch-of-intersection}
and thus separates the $(2d)$-dimensional phase space.
It also is useful for what follows to associate one degree of freedom 
$x_\mathrm{reac}$ as the reaction coordinate, and the 
remaining coordinates $\vec{x}_\mathrm{bath}$ as bath coordinates,
as indicated by the subscripts.
With these defined coordinates, for example, we can illustrate
the parameterized hypersurfaces such as is shown in 
\FIG~\ref{fig:network} (b).

Consideration of the barrier top dynamics in terms of the associated manifolds 
has the advantage that our geometric picture remains valid when the 
barrier is time-dependent.
The most important difference in the time-independent case is that the 
position at which the manifolds intersect is not identical with the 
instantaneous barrier top.
The manifold description also has the advantage that
it lends itself to calculation through 
the machinery of dynamical systems theory
including, for example, the very general method (cited in the introduction)
minimizing the LD,
\begin{equation}
\LD ( \vec x_0, \vec v_0, t_0) = 
\int_{t_0-\tau}^{t_0+\tau} \| \vec v(t) \| \, \mathrm{d}t \,.
\label{eq:LD}
\end{equation}
Due to the extremal property of trajectories on manifolds, the stable and 
unstable manifolds are related to the LD via
\begin{subequations}
\begin{align}
\mathcal{W}_\text{s} (t_0) &= 
\argmin \mathcal{L}^{(f)} (\vec x_0, \vec v_0, t_0) \,, \\
\mathcal{W}_\text{u} (t_0) &= 
\argmin \mathcal{L}^{(b)} (\vec x_0, \vec v_0, t_0) \,,
\end{align}%
\label{eq:Wsu}%
\end{subequations}
where ($f,b$) denote the forward ($f$: $t_0\leq t \leq t_0+\tau$) or the
backward ($b$: $t_0-\tau \leq t \leq t_0$) direction of time.
Here, the operator $\argmin$ denotes the argument of the local minimum. 
The intersection of both manifolds including their closure
\begin{equation}
\TSt(t_0) 
= \Ws(t_0) \cap \Wu (t_0)
\label{eq:LD-min}
\end{equation}
yields the NHIM at a given time $t_0$.

An essential challenge of exact reaction rate calculations is 
the tracking of a potentially
cost-prohibitively large number of particles in the ensemble, 
and the determination of the exact time of each crossing through the DS.
Further challenges when calculating such DSs are 
the numerical determination of the intersection of the manifolds
and the high dimensionality $2d-2$ of the NHIM, 
to which the DS is attached, as $d$ becomes large.
For a system with $d=1$, the calculation is trivial because it is a
single point.
The case $d=2$ is significantly more demanding leading to a two-dimensional 
surface embedded in full phase space which is numerically still manageable
using \eg~cubic spline interpolation.
For even higher dimensions, the effort to approximate the hypersurface, 
however, quickly becomes so large that following reactive particles becomes 
numerically intractable.
With increasing dimensionality, one needs a method that
interpolates smoothly and continuously across the surface.
Such an interpolation is necessary because the reactive particles
may cross the DS at an arbitrary phase space point 
for each initially unknown time.

It is the purpose of the remaining part of this paper to show that 
NNs are a 
useful tool for overcoming this dimensionality issue,
and that they can make numerical 
treatments possible for systems with a high number of bath degrees of freedom.

\subsection{Feed-forward neural networks}
\label{sec:theory:neural_nets_general}
%
\begin{figure}[t]
\includegraphics[width=\columnwidth]{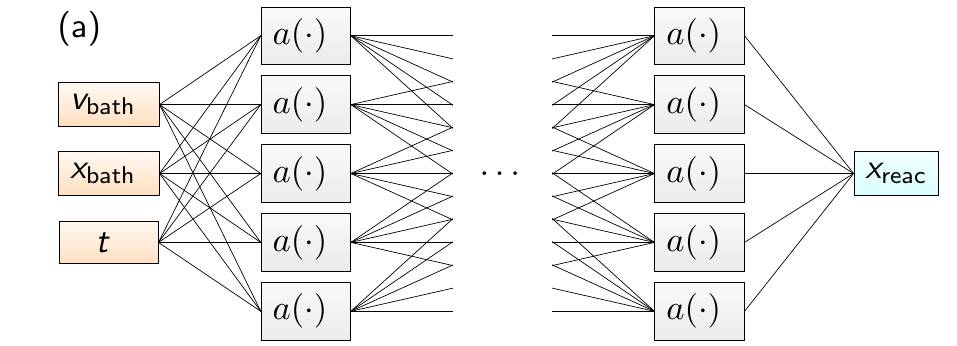}
\includegraphics[width=\columnwidth]{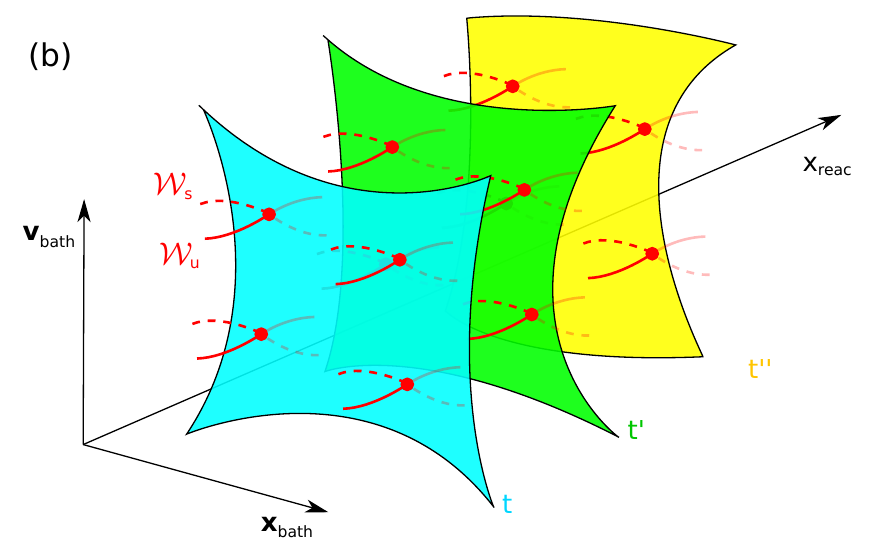}
\caption{%
(a) 
General structure of the neuronal network: 
Time $t$, the bath coordinates $\xx\subbath$, and their velocities $\vv\subbath$ define the input 
layer (orange nodes) and the reaction coordinate $x\subreac$ is the output layer 
(cyan node).
The gray nodes denote the hidden layers of adjustable depth and number of 
neurons, and $a(\cdot)$ is the respective activation function.
(b) 
General scheme of the procedure:
The time-dependent NHIM is defined by the intersection between the stable and 
unstable manifolds $\Wsu$ (solid and dashed red lines) for the respective, 
fixed values of the bath coordinates and for given time $t$.
The phase space coordinates of such an intersection point serve as the 
training 
set of the neural net.
}
\label{fig:network}
\end{figure}

In this section, we give a brief and basic overview of feed-forward NNs
to set the stage for our implementation to construct the 
time-dependent DS.
For details and more sophisticated types of NNs like recurrent 
or convolutional ones, we refer the reader to the extensive literature on this 
topic, such as \eg~\REFS~\cite{Goodfellow-et-al-2016,Nielsen-2015} and 
references therein.

In a feed-forward NN, information is processed 
consecutively by propagating it through a number of layers, each of which 
performs a nonlinear transformation of the input(s) resulting in
the output(s).
Such NNs typically consist of one input layer, one output layer as well as one 
or several in-between hidden layers, 
and each layer is made of several neurons, as 
illustrated in \FIG~\ref{fig:network}.
Each neuron in the hidden layers is connected to the neurons of the previous 
and the subsequent layer.

The most basic component, the neuron, is responsible for the information 
processing which works as follows:
The input $\Sigma$ of a neuron $i$ in layer $l$ is
a function of the information ---that is, outputs $a_{j}^{(l-1)}$---
from the previous layer $(l-1)$
and is given by
\begin{equation}
  \Sigma_i^{(l)} = \left(\sum_{j=1}^N w^{(l)}_{ij} a_j^{(l-1)} \right) + b_i^{(l)}\,.
  \label{eq:NN-weighted-input-def}
\end{equation}
The weight $w^{(l)}_{ij}$
represents the connection between the neurons
and reflect the importance of the information from the 
the previous neuron $j$ to the neuron $i$.
The value $ b_i^{(l)}$ is a threshold which determines the 
area of activation and is also referred to as the bias.
Of great importance for the functionality and learning behavior of the NN 
are the activation functions 
\begin{equation}
	a^{(l)}_i = a\left(\Sigma_i^{(l)}\right)\,.
	\label{eq:NN-activation-function}  
\end{equation}
They are generally nonlinear and their choice depends on the specific task 
of the NN.
In this paper, we use the inverse tangent $a(\Sigma)=\arctan(\Sigma)$
as activation function for the hidden layers.
For the input and output layer we use the identity matrix as the
weight of the activation function.

The weights and biases in \EQ~\eqref{eq:NN-weighted-input-def} serve as free 
parameters which are, during the network training, adjusted such that the NN 
performs the desired task.
The intended behavior is trained by providing a cost function $C$ 
which acts as a penalty term defining the network's deviation between its 
actual output $\tilde{\vec{y}}$ and the desired result $\vec{y}$.
To train the neural network, the cost function is minimized,
\eg~using 
back propagation and standard numerical procedures such as (batch) stochastic 
gradient descent.
In this paper, we use the standard mean squared error cost function
\begin{align}
	C_{\vec w, \vec b}(\vec{y},\tilde{\vec{y}}) = 
\frac{1}{2n}\sum\limits_{i=1}^{n}\left\lVert \vec{y}_i - 
\tilde{\vec{y}}(\vec{x}_i)\right\rVert^2\,,
	\label{eq:NN-cost-function-example}
\end{align}
where $\vec{x}_i$ is an input vector of the NN, $\vec{y}_i$ is the expected output, and $n$ 
is the total number of training points or the batch size, respectively. 
The function $\tilde{\vec{y}}(\vec{x}_i)$ is the output of the NN for the given 
input $\vec{x}_i$ and the subscripts ${\vec w, \vec b}$ denote the dependence of the 
cost function on the network parameters.

Using gradient descent, the weights and biases are updated in each step 
according to
\begin{subequations}
\begin{align}
w^{(l)}_{ij} &\rightarrow 
w^{(l)}_{ij} -\eta
\frac{\partial C_{\vec w, \vec b}}{\partial w^{(l)}_{ij}}\,,\\
b^{(l)}_{i} &\rightarrow 
b^{(l)}_{i} -\eta
\frac{\partial C_{\vec w, \vec b}}{\partial b^{(l)}_{i}}\,,
\label{eq:NN-gradient-descend}
\end{align}
\end{subequations}
where $\eta$ is the learning rate determining the step size of the gradient 
descent method.

The NN as described above was sufficiently simple to be implemented
without using special software packages.
However, use of freely available NNs will allow for more efficient
implementations and optimization of the training in the future.

\subsection{Application of neural networks to time-dependent dividing surfaces}
\label{sec:theory:neural_nets__applied_to_DSs}
In this paper, our goal is to use NNs for a multidimensional regression task 
in order to approximate the phase space DS of a reactive system.
The construction of the DS, compare \FIG~\ref{fig:sketch-of-intersection},
allows us to approximate the DS by only knowing the reaction coordinate of the NHIM for
a given set of bath coordinates and a given time.
More specifically, we want to learn the function \cite{FUNAHASHI1989183} that 
maps the current configuration of the system, \ie~the values of the bath degrees 
of freedom 
$\vec{x}_\mathrm{bath}, \vec{v}_\mathrm{bath}$ and time $t$, to the value 
${x}_\mathrm{reac}$ of the respective reaction coordinate,
\begin{align}
	{x}_\mathrm{reac} = 
	{x}_\mathrm{reac}(\vec{x}_\mathrm{bath},\vec{v}_\mathrm{bath},t) \,.
	\label{eq:NN-function-fit}
\end{align}
This is, in general, a complicated, time-dependent and nonlinear 
relation, but a finite number of representatives can be generated
readily using the method described in 
\SEC~\ref{sec:theory:LDs} through
minimization of the LD~\eqref{eq:LD-min}.
NNs are an ideal approach for extending the information from
this set of points---viz. the training set---to
provide an output for the entire domain.
Specifically, the NN is trained taking time $t$ and bath coordinates
$\vec{x}_\mathrm{bath}, \vec{v}_\mathrm{bath}$ as inputs to the
first layer, that leads to an input dimension of $2d-1$,
and the 
corresponding reaction coordinate ${x}_\mathrm{reac}$ as the outputs
from the last layer.

\subsection{Model system}
\label{sec:model-system}

To demonstrate the ability of NNs to approximate DSs,
we use a higher-dimensional extension of the 
model reactive system already investigated in 
\REF~\cite{hern17h}:
\begin{align}
V\left(x,y,z,t\right) =  \, E_\mathrm{b} 
\exp\left(-a_\mathrm{b}\left[x-x_\mathrm{b}\left(t\right)\right]^2\right)\nonumber\\
+ \frac{\omega_{y}^2}{2}\left[y-y_\mathrm{min}\left(x\right)\right]^2
+ \frac{\omega_{z}^2}{2}\left[z-z_\mathrm{min}\left(x\right)\right]^2\,.
\label{eq:model_pot}
\end{align}
This potential contains a Gaussian barrier along the $x$-direction
(which acts as $x_{\rm reac}$), where 
$E_\mathrm{b}$ is the barrier height and $a_\mathrm{b}$ the width. 
The orthogonal degrees of freedom, $y$ and $z$, act as the
bath coordinates ${\vec x}_\mathrm{bath}$.
The barrier oscillates along the $x$-axis according to
\begin{align}
x_\mathrm{b}\left(t\right) = x_{\mathrm{b},0}  \sin 
\left(\omega_{x_\mathrm{b}}t\right)\,,
\end{align}
where $x_{\mathrm{b},0}$ is the amplitude and $\omega_{x_\mathrm{b}}$ the 
frequency of the oscillation.
Further, the potential includes harmonic potentials along the $y$- and $z$-axis 
with the frequencies $\omega_{y,z}$ and the degrees of freedom are 
nonlinearly coupled to the $x$-direction along the respective minimum energy 
valleys
\begin{align}
y_\mathrm{min}\left(x\right)= z_\mathrm{min}\left(x\right)= 
\frac{2}{\pi}\arctan\left(2x\right)\,.
\end{align}
For simplicity, all variables are introduced 
in dimensionless, and are set to 
$E_\mathrm{b}=2$, $a_\mathrm{b}=1$, $\omega_y = 2$, $\omega_z = 1$, $x_{\mathrm{b},0} = 0.4$ and $\omega_{x_{\mathrm{b}}} = \pi$.
Due to the time dependence of the saddle,
the resulting DS is also be time-dependent.
Note that the two-dimensional model system previously discussed 
in \REF~\cite{hern17h} is obtained by neglecting the $z$-degree of freedom in \EQ~\eqref{eq:model_pot}.

\section{Results and Discussion}
\label{sec:results}

In this section, we demonstrate several aspects of the use of
NNs for obtaining DSs.
The starting point for all of the implementations
is a set of phase space points located 
on the NHIM that are precalculated according to \EQ~\eqref{eq:LD-min} for 
the respective potential~\eqref{eq:model_pot}.
See \FIG~\ref{fig:network}(b) for an illustration. 
Note that it is not within the 
scope of this paper to illustrate and explain the performed 
minimization process using \EQ~\eqref{eq:LD-min}
as that has already been confirmed and discussed in our recent
work \cite{hern17h}.

\begin{figure}[t]
\includegraphics[width=\columnwidth]{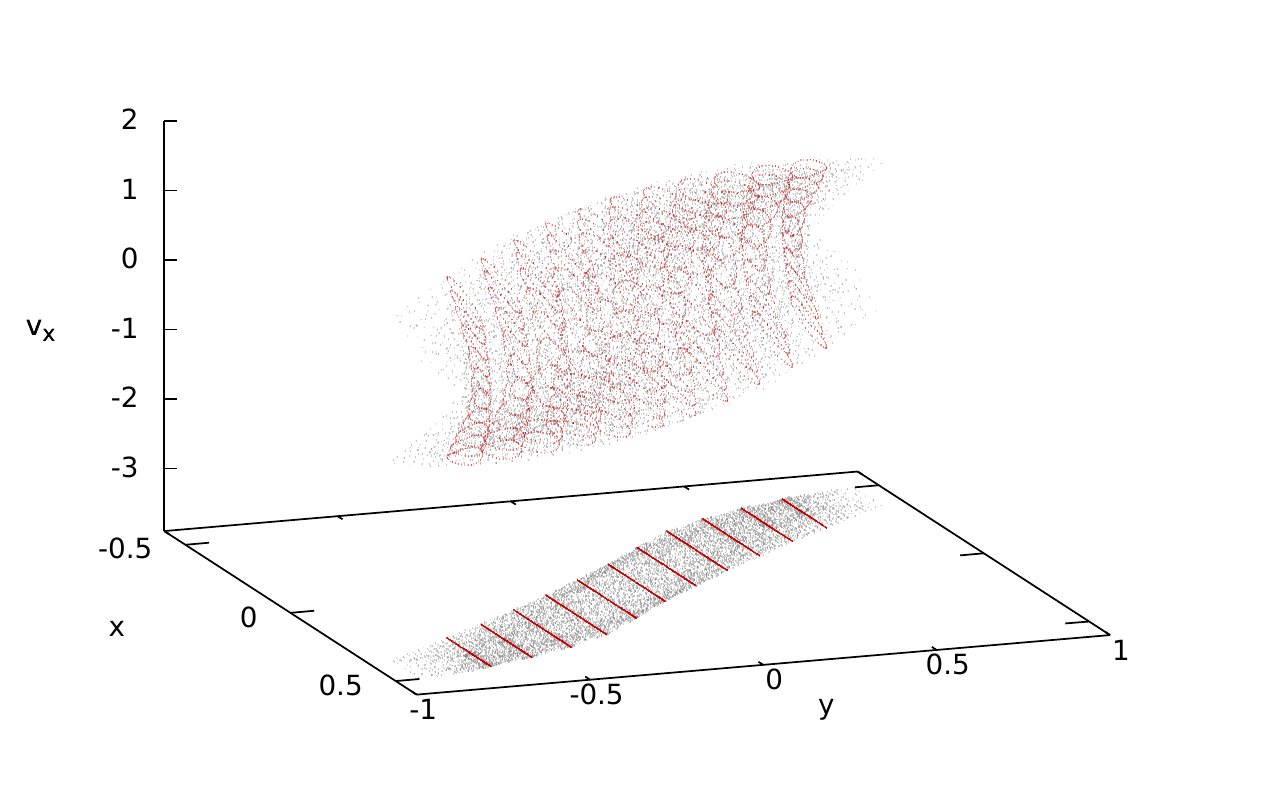}
\caption{%
Extension of the training set in 2D.
The red points are calculated using the LD minimization procedure according to 
\EQ~\eqref{eq:LD-min}.
The gray points are afterwards obtained by using the red dots as initial values 
and propagating the respective particle according to the equations of motion.
By this procedure the number of training points for the neural network is easily 
extended.
}
\label{fig:2D-extend-training-set}
\end{figure}

\begin{figure*}[t]
\includegraphics[width=0.95\textwidth]{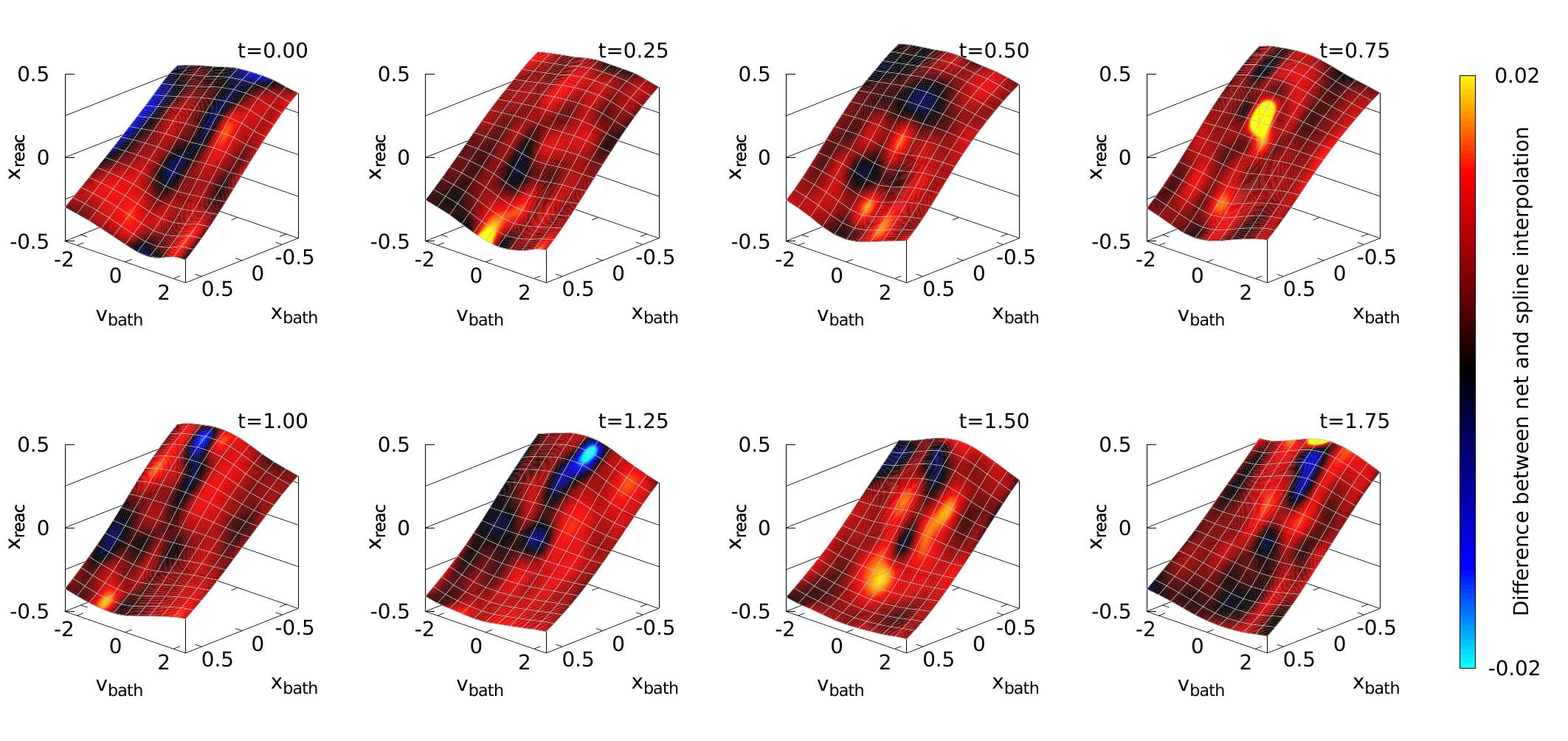}
\caption{%
Time-dependent NHIM obtained from the NN in dependence of the bath coordinates.
Note that $x_\mathrm{reac}$ refers to $x$ and $x_\mathrm{bath}$ and $v_\mathrm{bath}$ to $y$ and its velocity, respectively, in \EQ~\eqref{eq:model_pot}.
The color of the surface indicates the difference between the NN's 
output and a spline interpolation of the data on which the NN is trained.
Here, the smoothing ability of the neural network becomes visible.
}
\label{fig:2D-surfaces-timedependent}
\end{figure*}

To obtain the training set, 
we perform the minimization~\eqref{eq:LD-min} only for a limited 
number of points---\eg~on an equidistant grid in the bath coordinates and in 
time---highlighted in red in \FIG~\ref{fig:2D-extend-training-set}.
Since all these points are, by construction, located on an invariant manifold, 
this structure remains valid as the respective particle is propagated in time.
Thus, we can readily add new points to the training set as needed.
For example, we can extend the training set size 
by a factor of ten 
simply by integrating five time steps in forward and backward 
time (if only a few steps are performed, there is no need to take special care 
of the instability of the trajectory).
As a consequence, the newly calculated points are completely 
unordered but nevertheless associated with the appropriate corresponding time.
Consequently, their use in training the NN,
although formally out of order, introduces no error.

\subsection{System with two degrees of freedom}\label{sec:results:2dof}

First, we apply the NN to approximate a time-dependent DS for a system with two 
degrees of freedom.
In this case, we neglect the $z$-degree of freedom in \EQ~\eqref{eq:model_pot}, 
and we use a NN with two hidden layers consisting of $40$ and $10$ neurons.
Further, the NN is trained on 2400 points of the NHIM and the training
is stopped after $10^6$ training cycles.
Figure~\ref{fig:2D-surfaces-timedependent} shows the result of the NN 
regression: snapshots of the NHIM at different times over one period.
Specifically, the reaction coordinate $x$ is shown over the 
domain of bath coordinates $y$ and $v_y$ ($v_x$ is not shown).
The surfaces displayed are the actual output of the neural network and the 
color code refers to the deviation compared to a spline interpolation of the 
underlying training set.
The deviation between the training data points and the neural network prediction is 
small throughout.
Its accuracy is also confirmed by the small value of 
the mean squared training error (=$2.95 \times 10^{-6}$) that we 
achieved at the end of the training procedure.

At some times and grid points, small areas show large deviations between the 
neural network output and the spline interpolation (cyan and yellow spots on 
the surface).
However, as we have verified by a direct comparison with training data, these 
points are not erroneous predictions of the NN, 
but rather indicate grid points where the 
original training point has been poorly determined during the minimization 
procedure~\eqref{eq:LD-min}.
The NN is thus smoothing out such numerical errors in the training set due 
to its global learning property 
(whereas spline methods just interpolate between points).
This is another advantage for the use of NNs to numerically approximate DSs.

\subsection{System with three degrees of freedom}

\begin{figure}
\includegraphics[width=0.9\columnwidth]{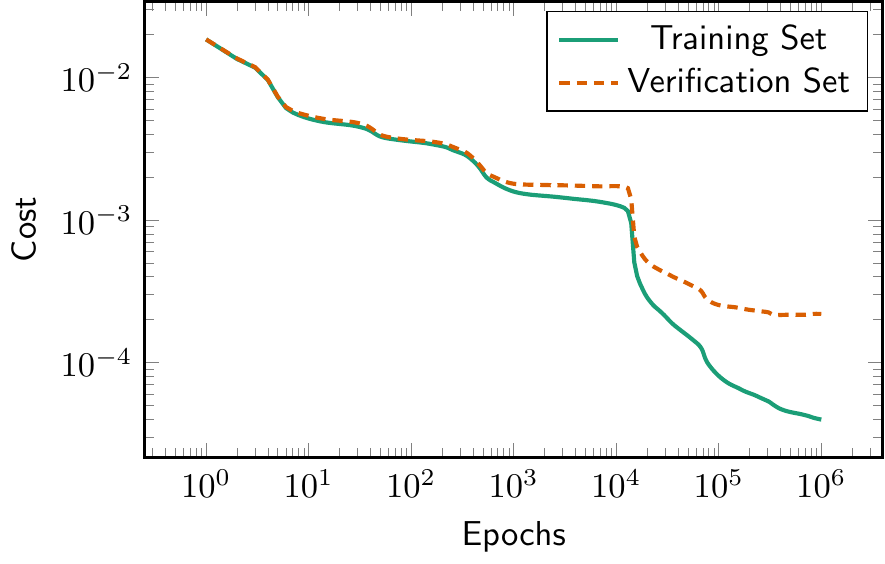}
\caption{
Cost functions of the data set for a DS of the system with three degrees of 
freedom.
The cost function represents the difference between the output of the neural 
network and the actual points in the data set.
The training set is a subset of the whole data set on which the neural network 
is trained and the verification set is a disjunct second subset on which the 
neural network is not trained.
So the cost function of the verification set
is a measure of how accurately the neural network 
interpolates.
The training was stopped after $10^6$ epochs.
}
\label{fig:cost_func_3d}
\end{figure}

We now take into account the full system~\eqref{eq:model_pot} with three 
degrees of freedom, and calculate $10,000$ points of the NHIM as our training 
and verification sets ($4,000$ and $6,000$ points, respectively).
The input layer of the NN now requires five neurons for all four bath 
coordinates, $y,z,v_y,v_z$, and one neuron for the time dependence of the NHIM.
Here, we use two hidden layers with $40$ and $10$ neurons, respectively.
The NN parameters are updated through each full loop --- called an epoch ---
of the training algorithm across the complete training data set.
The cost function of the evolving NN in terms of epochs 
shows the training progress in \FIG~\ref{fig:cost_func_3d}.
Note that the cost for the training and verification sets nearly agree
up to the first $10^3$ epochs.
For more epochs, the cost function of the verification set decreases less
rapidly, indicating a saturation of the interpolation accuracy
as is typical for NNs.

\begin{figure}
\includegraphics[width=0.9\columnwidth]{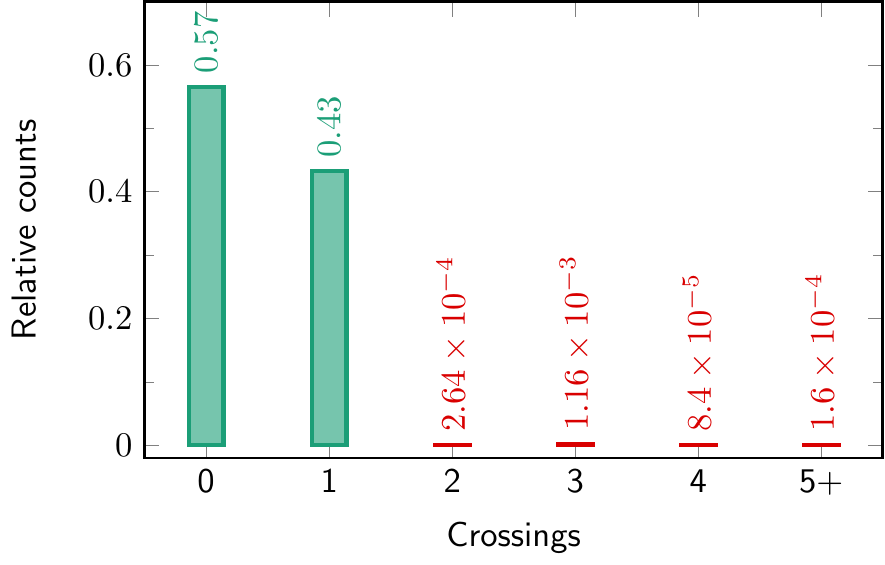}
\caption{
Counted crossings through the DS approximated by a neural network.
The corresponding reaction curve can be seen in 
\FIG~\ref{fig:reaction_curve_3d_system}.
The bars corresponding to two and more crossings are recrossings.
Taking into account the small number of recrossings, the approximated DS is a 
recrossing-free DS with only tiny numerical errors.
}
\label{fig:crossings_3d}
\end{figure}

In order to verify the recrossing-free property of the DS generated by the 
NN, we regard a thermal ensemble of $250,000$ particles in the reactant well 
($x<0$) with a density distribution
\begin{align}
	\rho \left(\vec{x},\vec{v}\right) = \rho_\mathrm{therm}\delta\left(x+1\right)\Theta\left(v_x\right)\,,
\end{align}
where $\rho_\mathrm{therm}$ is a Boltzmann distribution, $\delta$ is the Dirac 
delta function and $\Theta$ is the Heaviside step function.
We use these functions because the potential is unbound at
$x\rightarrow\pm\infty$, and particles with outward velocities would not 
take part in the reaction.
The particles of the generated thermal ensemble are propagated and any crossing 
through the DS is counted. 

The relative counts of crossings are shown in \FIG~\ref{fig:crossings_3d}.
The two green bars for zero and one crossing do not violate the recrossing-free 
property of an exact DS 
while the red bars ---for multiple crossings--- violate it.
Almost all particles either show no or one crossing, and only a negligible 
fraction ($1.67 \times 10^{-3}$) of all particles show recrossings at all.
Thus, the recrossing-free property is fulfilled to a high degree.

\begin{figure}
\includegraphics[width=0.9\columnwidth]{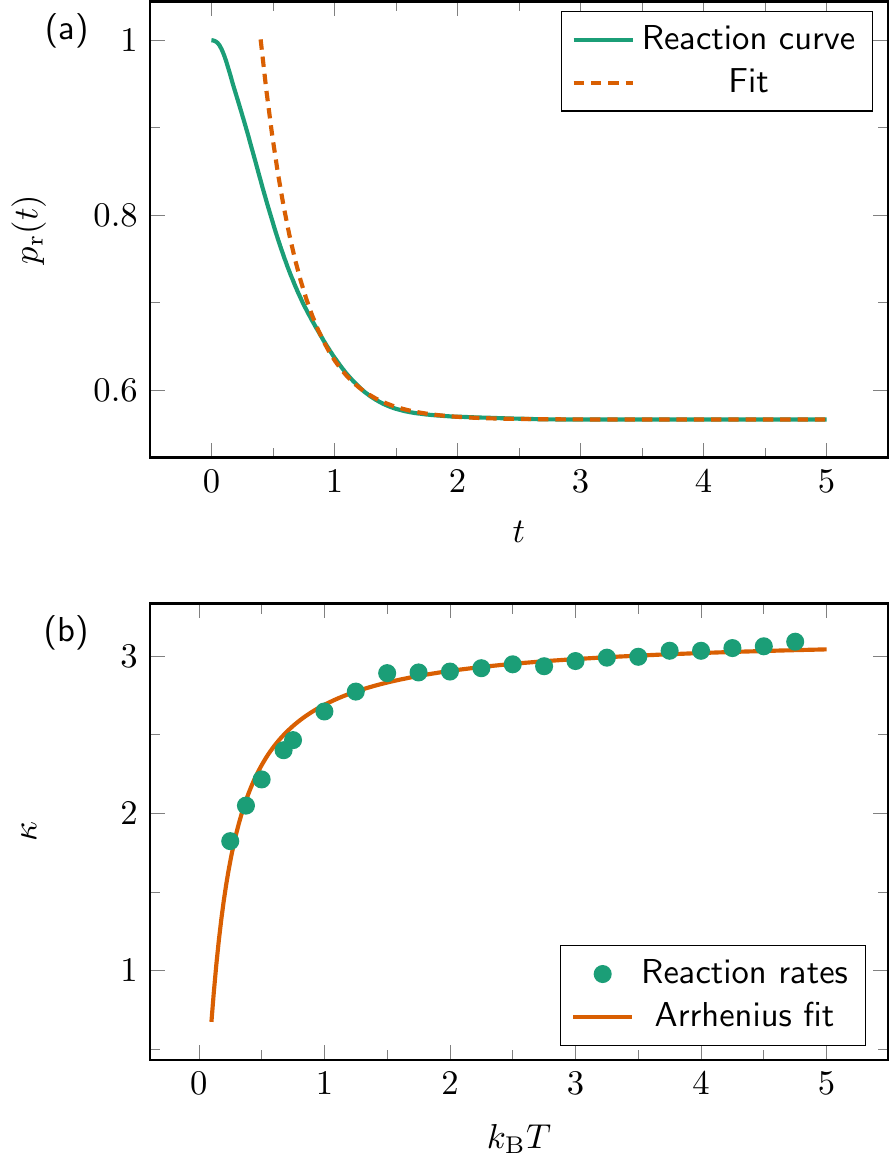}
\caption{
(a) Behavior of the reactant population over time of a thermal ensemble at 
$k_\mathrm{B}T=4.75$ in units of $E_\mathrm{b}/2$.
The population is normalized to the amount of particles propagated, which were 
$250,000$ particles.
There is also an exponential fit (dashed line) to this reaction curve as in 
\EQ~\eqref{eq:exp-fit} which yields a reaction rate of $\kappa = 3.09232$.
(b) Reaction rates determined with fits as in subfigure (a) for different values 
of $k_\mathrm{B}T$.
The fit corresponds to the Arrhenius equation \eqref{eq:arrhenius-equation}.
}
\label{fig:reaction_curve_3d_system}
\end{figure}

The corresponding population 
decay of the initial ensemble is shown
in \FIG~\ref{fig:reaction_curve_3d_system}(a),
The decay exhibits an expected monotonous decrease and the subsequent 
saturation to a constant population. 
The dashed line is an exponential fit
\begin{align}
	p_{\textrm{r}}(t) = p_{{\textrm{r}},0} \ue^{-\kappa t} + p_{\textrm{r},\textrm{c}}
	\label{eq:exp-fit}
\end{align}
to the long-time decay of the reactant population with the fit parameters 
$p_{{\textrm{r}},0}$, $\kappa$ and $p_{\textrm{r},\textrm{c}}$, where $\kappa$ 
corresponds to the reaction rate.
The population decay agrees well with the exponential fit as expected 
for a first-order transition.
The reaction rate $\kappa$ is further shown for ensembles of different 
temperature $k_\mathrm{B}T$ in \FIG~\ref{fig:reaction_curve_3d_system}(b) as 
green dots.
The solid line refers to the Arrhenius' rate equation
\begin{align}
	\kappa(k_\mathrm{B}T) = \kappa_\infty \ue^{- \Delta E_\mathrm{eff}/ k_\mathrm{B}T}\,,
	\label{eq:arrhenius-equation}
\end{align}
where $\kappa_ \infty$ is the high-temperature limit of the reaction rate and 
$E_\mathrm{eff}$ denotes the effective height of the energy barrier.
We obtain $\kappa_ \infty = 3.138$ and $E_\mathrm{eff}=0.154$ for the fit parameters.

\section{Conclusion}

In this paper, we have introduced neural networks as a tool for
approximating high-dimensional, 
time-dependent DSs in classical reaction dynamics.
We have shown that with rather small effort in calculating points on the NHIM
as the training set for a NN, the resulting network predicts the
DS accurately over a broader domain along a continuous times interval.
This holds true even in the more challenging case of increased dimensionality
of the bath.
The accuracy and exactness of the resulting DS reproduced by the NN 
has been verified through the observation of
negligible recrossings and calculated rates that are in
good agreement with theory.
We anticipate that DSs can be obtained using neural networks in systems of even
more degrees of freedom, 
although such an approach is clearly more challenging as the 
dimensionality grows beyond the three evidenced here.

\section*{Acknowledgment}
The German portion of this collaborative work was partially supported by DFG.
The US portion was partially supported by the 
National Science Foundation (NSF) through Grant No.~CHE 1700749.
AJ acknowledges the Alexander von Humboldt Foundation, Germany, for support 
through a Feodor Lynen Fellowship.
MF is grateful for support from the Landesgraduiertenf\"orderung of
the Land Baden-W\"urttemberg.
This collaboration has also benefited from support by the 
European Union's Horizon 2020 Research and Innovation Program
under the Marie Sklodowska-Curie Grant Agreement
No.~734557.


\bibliography{paperq14}

\end{document}